\documentclass[12pt,aps,prb,preprint]{revtex4}   
\usepackage{amsmath}    
\usepackage{graphicx}   
\begin{document}

\title{Mie plasmons: modes volumes, quality factors and coupling strengths (Purcell factor) to a dipolar emitter.}
\author{G. {Colas des Francs}}
\email{gerard.colas-des-francs@u-bourgogne.fr}
\author{S. Derom}
\author{R. Vincent}
\author{A. Bouhelier}
\author{A. Dereux}

 \affiliation{Laboratoire Interdisciplinaire Carnot de Bourgogne, UMR 5209 CNRS - Universit\'e de Bourgogne, \\
9 Av. A. Savary, BP 47 870, 21078 Dijon, FRANCE}  

\date{\today}

\newcommand{\gras}[1]{\mathbf{#1}}
\begin{abstract}
Using either quasi-static approximation or exact Mie expansion, we characterize the localized surface plasmons supported by a metallic spherical nanoparticle. 
We estimate the quality factor $Q_n$ and define the effective volume $V_n$ of the $n^{th}$ mode in a such a way that coupling strength with a neighbouring dipolar emitter is 
proportional to the ratio $Q_n/V_n$ (Purcell factor). The role of Joule losses, far-field scattering and mode confinement in the coupling mechanism are introduced and discussed with simple physical understanding, with particular attention paid to energy conservation.
\end{abstract}

\maketitle

\section{Introduction}
Metallic nanoparticles support localized surface plasmon-polaritons (SPP) strongly confined at the metal surface ensuring efficient 
electromagnetic coupling with neighbouring materials, offering a variety of applications such as 
surface-enhanced spectroscopies \cite{Moskovits:1985,Giannini-FrenadezDominguez-Heck-Maier:2011}, 
photochemistry \cite{Deeb-Bachelot-Plain-Soppera:2010} or optical nanoantennas \cite{Bharadwaj-Deutsch-Novotny:2009}. 
This also opens the way towards control of light emission at the 
nanoscale \cite{Bergman-Stockman:2003,Schietinger-Barth-Aichele-Benson:2009,Huang-BerthelotAPL:2010,Marty-Arbouet-Paillard-Girard-GCF:2010,Vincent-Carminati:2011,Hyung-Park:2011}. 
The ratio $Q/V_{eff}$ is generally used to quantify the coupling strength between a dipolar emitter and a cavity mode (polariton). 
$Q$ and $V_{eff}$ refers to the mode quality factor and effective volume, respectively. 
High ratio $Q/V_{eff}$ may lead to strong coupling regime with characteristics Rabi oscillations revealing cycles of energy exchange 
between the emitter and the cavity. 
In the weak coupling regime, the emitter energy dissipation into the cavity mode is non reversible and follows the Purcell factor:
\begin{equation}
\frac{\gamma}{n_B \gamma_0}=\frac{3}{4\pi^2}\left(\frac{\lambda}{n_B} \right) ^3 \frac{Q}{V_{eff}} \,,
\label{eq:Purcell}
\end{equation}
where $\gamma$ is the emitter spontaneous decay rate into the cavity compared to its free-space value $\gamma_0$, $n_B$ is the optical index inside the cavity and $\lambda$ is the emission wavelength. Spontaneous emission rate in complex systems follows the more general Fermi's golden rule that expresses $\gamma$ as a function of the density of electromagnetic mode \cite{CPLGirardGCF:2005}.  However, for describing the coupling to a cavity mode, the Purcell factor is usually preferred since it clearly introduces the cavity resonance quality factor $Q$ and the mode extension $V_{eff}$ on which coupling remains efficient, bringing therefore a clear physical understanding of the coupling process. 

In this context, we propose to determine the quality factors and effective volumes of localized SPPs supported by a metallic nanosphere. 
Indeed, these quantities are usefull parameters to understand and evaluate the coupling mechanisms between dipolar emitters 
and plasmonic nanostructures \cite{Maier:2006,Datsyuk:2007}. This would help for achieving strong coupling regime \cite{Truegler-Hohenester:2008,Savasta-Borghese:2010} or for designing plasmonic nanolasers \cite{Protsenko:2005,NoginovNature:2009,Stockman:2010}. We have chosen a spherical particle, a highly symetrical system, since it is fully analytical and simple expressions can be derived with clear physical meaning \cite{JCPGCF:2005,Carminati-Greffet-Henkel-Vigoureux:2006,OptExpGCF:2008,GCFIJMS:2009,Rolly-Stout-Bidault-Bonod:2011}. Moreover, our results could be extended to more complex structures \cite{Bonod-Devilez-Rolly-Bidault-Stout:2010,Sun-Soref:2011}.

In section \ref{sect:dipole}, we focus on the dipolar mode and present in details the derivation of its quality factor and effective volume. 
We extend our approach to each mode of the particle in section \ref{sect:multipole}. Finally, we discuss the coupling efficiency to one of the particle modes in the last section. 
For the sake of simplicity, all the analytical expressions are derived assuming a particle in air and dipolar emitter perpendicular to the particle surface. The generalisation to arbitrary emitter orientation and a background medium of optical index $n_B$ is provided in the appendix. Exact calculations presented in the main text using Mie expansion correctly include both the background medium and dipolar orientation.

\section{Dipolar mode}
\label{sect:dipole}
We first characterize the dipolar mode of a spherical particle. For sphere radius $R$ small compared to the excitation wavelength $\lambda=2\pi c/\omega$, the electric field is considered uniform over the metallic particle. The metallic particle is polarized by the incident electric field ${\bf E_0}$ and behaves as a dipole 
\begin{eqnarray}
{\bf p}^{(1)}(\omega)=4 \pi \epsilon_0 \alpha_1(\omega) {\bf E_0} \,, \\ 
\alpha_1(\omega)=\frac{\epsilon_m(\omega)-1}{\epsilon_m(\omega)+2} R^3 \,,
\label{eq:alpha_1}
\end{eqnarray} 
where $\alpha_1$ is the nanoparticle quasi-static (dipolar) polarisability and $\epsilon_m$ is the metal dielectric constant. The dipole plasmon resonance appears at $\omega_1$ such that $\epsilon_m(\omega_1)+2=0$. In case of Drude metal, the dipolar resonance is $\omega_1=\omega_p/\sqrt{3}$ with $\omega_p$ the bulk metal plasma angular frequency. However, expression (\ref{eq:alpha_1}) does not satisfy the optical theorem (energy conservation). It is well-known that this apparent paradox is easily overcome by taking into account the finite size of the particle and leads to define the effective polarisability \cite{Wokaun-Gordon-Liao:1983}:
\begin{equation}
\label{eq:alpha_eff1}
\alpha_1^{eff}=\left[ 1 - i~\frac{2k^{3}}{3}\alpha_1 \right] ^{-1} \alpha_1 \;\;, (k=2\pi / \lambda) \,. 
\end{equation} 
The corrective term ($2k^3 \alpha_1/3$) is the so-called radiative reaction correction and microscopically originates from the radiation emitted by the charges oscillations induced inside the nanoparticle by the excitation field \cite{Jackson:1998}.
\par
\bigskip
The dipolar polarisability presents a simple shape near the resonance if the metallic dielectric constant follows Drude model \cite{Carminati-Greffet-Henkel-Vigoureux:2006} ($\omega_p$ and $\Gamma_{abs}$ refers to metal plasma frequency and Ohmic loss rate, respectively) \cite{Notation}; 
\begin{eqnarray}
\epsilon_m&=&1-\frac{\omega_p^2}{\omega^2+i\Gamma_{abs} \omega} \,;\\
\label{eq:alpha1Drude}
\alpha_1^{eff}(\omega)&\underset{\omega_1}{\sim}&\frac{\omega_1}{2(\omega_1-\omega)-i\Gamma_1} R^{3}\,, \\
\Gamma_1&=&\Gamma_{abs}+\frac{2(k_1 R)^3 \omega_1}{3}  \,, (k_1=\omega_1/c) \,.
\end{eqnarray}
where $\Gamma_1$ is the decay rate of the particle dipolar mode, and includes both the Joule ($\Gamma_{abs}$) and radiative [$\Gamma_1^{rad}=2(k_1R)^3 \omega_1/3$] losses rates. 

\subsection{Quality factor}
The dipolar response can be described by either the extinction efficiency $Q_{ext}$, proportional to $Im(\alpha_1)$, or scattering efficiency $Q_{scatt}$, proportional to $|\alpha_1|^2$, and therefore follows a Lorentzian profile centered at $\omega_1$ and with a full width at half maximum (FWHM) $\Delta \omega_1=\Gamma_1$:
\begin{eqnarray}
Q_{ext}(\omega) &=& \frac{4k}{R^2} Im[\alpha_1^{eff}(\omega)] \\
\nonumber
&\underset{\omega_1}{\propto}&
\frac{1}{(\omega-\omega_1)^2+(\Gamma_1/2)^2}  \,, \\ 
Q_{scatt}(\omega) &=&\frac{8}{3R^2}k^4|\alpha_1^{eff}(\omega)|^2 \\
\nonumber
&\underset{\omega_1}{\propto}&
\frac{1} {(\omega-\omega_1)^2+(\Gamma_1/2)^2}  \,.
\end{eqnarray}
The quality factor of this resonance is therefore:
\begin{equation}
Q_{1}=\frac{\omega_1}{\Gamma_1} \,.
\end{equation}

\begin{figure}[h!]
\begin{center}
\includegraphics[width=8cm]{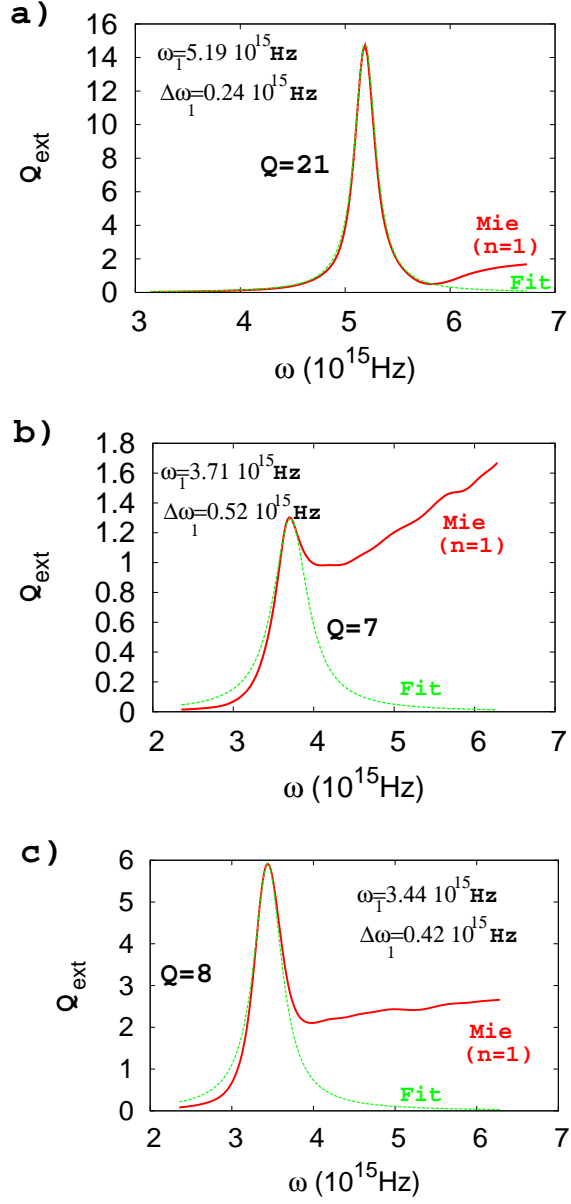}
\caption{\label{fig:Qdip} Extinction efficiency for the dipolar resonance calculated keeping only the dipolar mode ($n=1$) in the Mie expansion and using experimental value for the metal dielectric constant \cite{Johnson-Christy:1972}. a) Silver particle in air. b) Gold particle in air. c) Gold particle in PMMA (optical index $n_B=1.5$). The particle radius is $R=25~nm$ for each case. Fit refers to a Lorentzian fit using parameters indicated on the figure.}
\end{center}
\end{figure}  
As an example, we consider a $R=25~nm$ silver sphere in air. The Drude model parameters are $\hbar \omega_p=9.1~eV$ and  $\hbar \Gamma_{abs}=18~meV$ 
that lead to a resonance peak at $\hbar \omega_1=5.2~eV$ ($\omega_1=7.98~10^{15}~Hz$) and radiative energy $\hbar \Gamma_1^{rad}=1~eV$. 
The quality factor is then $Q_{1}=5$. Note that the metal optical properties are better described when including the bound electrons 
in the Drude model: $\epsilon_m=\epsilon_{\infty}-\omega_p^2/(\omega^2+i\Gamma_{abs} \omega)$, ($\epsilon_{\infty}=3.7$ for silver, see also the appendix). 
In that case, we obtain $\hbar \omega_1=3.74~eV$ ($\omega_1=5.7~10^{15}~Hz$), $\hbar\Gamma_1^{rad}=0.13~eV$ and $Q_{1}=24$.


Although Drude model qualitatively explains the shape of the resonance, a more representative value of the quality factor can only be determined using tabulated data for the dielectric constant of the metal \cite{Johnson-Christy:1972}. 
The extinction efficiencies associated to the dipolar resonance are represented in Fig. \ref{fig:Qdip}. For silver particle (Fig. \ref{fig:Qdip}a), the extinction efficiency closely follows a Lorentzian shape, 
as expected, with a quality factor $Q_{1}=21$, in good agreement with the value obtained using Drude model (with the contribution of the bound electrons included). 
In case of gold, the resonance profile is not well defined due to interband transitions for $\omega > 4.10^{15}~{Hz}$ 
but a quality factor of 7 can be estimated (Fig. \ref{fig:Qdip}b). 
Figure \ref{fig:Qdip}c) represents the extinction efficiency for a gold particle embedded in polymethylmethacrylate (PMMA) 
into which metallic particles are routinely dispersed. 
This leads to a small redshift of a resonance, avoiding therefore the resonance disturbance by interband absorption 
and we recover partly the Lorentzian profile \cite{sonnichsen02}.  
\subsection{Effective volume}
Coupling rate of a dipolar emitter to a dipolar particle expresses for very short emitter-particle distances $d$ ($z_0=R+d$ is the distance to the particle center) as \cite{JCPGCF:2005,Carminati-Greffet-Henkel-Vigoureux:2006,OptExpGCF:2008}:
 \begin{equation}
 \label{eq:gz}
 \frac{\gamma_{1}^{\perp}}{\gamma_0}\sim \frac{6}{k^3 z_0^6}
 Im(\alpha_1) \,,
 \end{equation} 
for a dipole emitter orientation perpendicular to the nanoparticle surface. Using Eq. \ref{eq:alpha1Drude}, we obtain, in case of a dipolar emitter emission tuned to the dipolar particle resonance ($\lambda=\lambda_1=2\pi c/\omega_1$)
 \begin{equation}
  \frac{\gamma_{1}^{\perp}}{\gamma_0}\underset{\omega_1}{\sim}\frac{6 \omega_1 R^3}{k_1^3 z_0^6 \Gamma_1}\sim \frac{3}{4\pi^2}\lambda_1 ^3
\frac{R^3}{\pi z_0^6} Q_{1} \,.
 \end{equation} 
In order to determine the dipolar mode effective volume, we now identify the coupling rate $\gamma_/\gamma_0$ to the Purcell factor (Eq. \ref{eq:Purcell}, assuming $n_B=1$), so that we obtain 
\begin{eqnarray}
V_{1}^{\perp}=\frac{\pi z_0^6}{R^3} \,,
\end{eqnarray}
in full agreement with expression recently derived by Greffet {\it et al} by defining 
the optical impedance of the nanoparticle antenna \cite{Greffet-Laroche-Marquier:2010}. 
For a $R=25 ~nm$ radius sphere in air, we estimate the mode effective volume $V_{1}^{\perp}=3.7~10^{-4}~\mu m^3=(72~nm)^3$ 
for an emitter 10 nm away from the particle surface ($z_0=35~nm$).

Unlike here, usual definition of the mode effective volume does not include the emitter position. 
For instance, in case of cavity quantum electrodynamics (cQED) applications, it can be expressed as 
$V_{eff}=\int \epsilon |{\bf E}({\bf r})|^2 d{\bf r} /max(\epsilon |{\bf E}({\bf r})|^2)$, so that it directly 
characterizes the mode extension. However, in that case, Purcell factor expression (Eq. \ref{eq:Purcell}) is only valid 
for an emitter located at the cavity center where the mode--emitter coupling is maximum (mode antinode). 
In close analogy with the definition used in cQED, we derive later another expression for the SPP effective volume, 
independant on the emitter position (see Eq. \ref{eq:Veff}, and the discussion below).

\section{Multipolar modes}
\label{sect:multipole}
\begin{figure}[h!]
\begin{center}
\includegraphics[width=16cm]{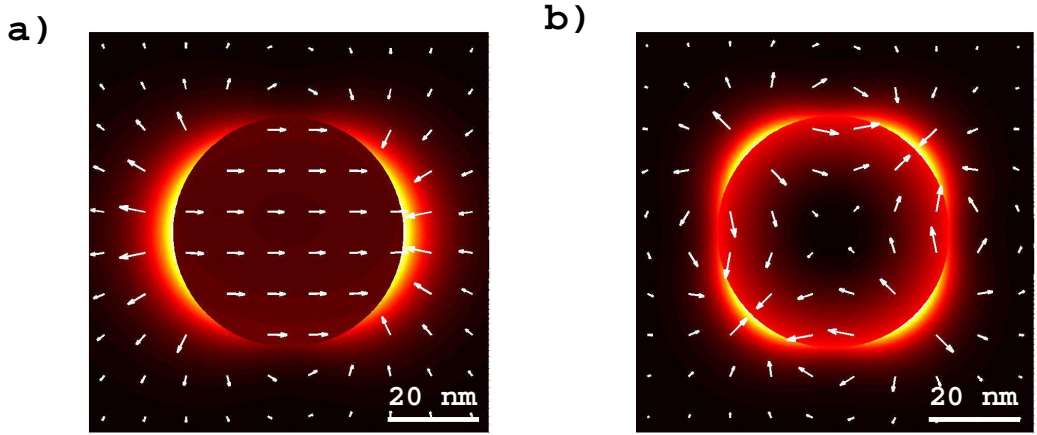}
\caption{\label{fig:AgR25inPMMA} a) Dipolar and b) quadrupolar mode profiles of a $R=25~nm$ silver sphere, embedded in PMMA calculated using exact Mie expansion. 
Silver dielectric constant is taken from ref. \cite{Johnson-Christy:1972}. Color and arrows refer to electric field intensity and vector respectively.}
\end{center}
\end{figure}  
If the excitation field is generated by a dipolar emitter (fluorescent molecules, quantum dots, \ldots), it cannot be considered uniform anymore and the dipolar approximation fails. 
One needs therefore to consider the coupling strength to high order modes (Fig. \ref{fig:AgR25inPMMA}). 
Here again, we discuss the mode quality factors and volume first using quasi-static approximation 
and then discuss their quantitative behaviour using exact Mie theory. 
 
The $n^{th}$ multipole tensor moment of the metallic particle is given by: 
\begin{eqnarray}
{\bf p}^{(n)}=\frac {4 \pi \epsilon_0} {(2n-1)!!} \alpha_n \nabla ^{n-1} {\bf E_0} \,,\\
\alpha_n=\frac{n(\epsilon_m-1)}{n\epsilon_m+(n+1)}R^{(2n+1)} \,,
\label{eq:alpha_n}
\end{eqnarray}
with $(2n+1)!!=1\times 3\times 5\times \ldots \times (2n+1)$ and $\nabla$ is the vector differential operator. As discussed above, the dipole moment ${\bf p}^{(1)}=4 \pi \epsilon_0 \alpha_1 {\bf E_0}$ is the unique mode excited in an uniform field. However, the dipolar emitter near-field behaves as $1/r^3$ and strongly varies spatially so that higher modes can be excited in the particle.

For a Drude metal, the $n^{th}$ resonance appears at $\omega_n=\omega_p \sqrt{n/(2n+1)}$. 
Therefore, higher order modes accumulate near $\omega_{\infty}=\omega_p/\sqrt2$. Moreover, as discussed for the dipolar case, quasi-static expression of the $n^{th}$ mode polarisability (Eq. \ref{eq:alpha_n}) does not obey energy conservation. Applying the optical theorem, we recently extend the radiative correction to all the spherical particles modes \cite{GCFIJMS:2009}. This leads to
\begin{equation}
 \label{eq:alpha_eff}
\alpha_n^{eff}=\left[1 - i~\frac{(n+1)k^{2n+1}}{n(2n-1)!!(2n+1)!!}\alpha_n \right ]^{-1}\alpha_n \,.
 \end{equation}

\subsection{Quality factors}
It is now a simple matter to generalize the dipolar mode analysis reported in the previous section to all the particle modes. The $n^{th}$ mode polarisablity can be approximated near resonance. A simple expression for the polarisability is achieved considering a Drude metal:
\begin{eqnarray}
\label{eq:alphaNapprox}
\alpha_n^{eff}&\underset{\omega_n}{\sim}&\frac{\omega_n}{2(\omega_n-\omega)-i\Gamma_n} R^{2n+1}\,,\\
\nonumber
\Gamma_n&=&\Gamma_{abs}+\Gamma_n^{rad} \,, \\
\nonumber
\Gamma_n^{rad}&=&\omega_n \frac{(n+1)(k_n R)^{2n+1}}{n(2n-1)!!(2n+1)!!} \,, (k_n=\omega_n/c) \,,
\end{eqnarray}
where $\Gamma_n$ is the total decay rate of the $n^{th}$ mode, that includes both ohmic losses and radiative scattering. 
As expected, for a given mode $n$, the radiative scattering rate $\Gamma_n^{rad}\propto R^{2n+1}$ increases with the particle size 
since it couples more efficiently to the far-field.  
For instance, we obtain $\hbar \Gamma_2^{rad}=39~meV$ ($\hbar \Gamma_2^{rad}=1.8~meV$ with $\epsilon_{\infty}=3.7$) 
for the quadrupolar mode of a $R=25~nm$ silver particle in air. 
As expected, the radiative rate of the quadrupolar is strongly reduced compared to the dipolar mode.

The quality factor associated to the $n^{th}$ mode is therefore 
\begin{equation}
Q_n=\frac{\omega_n}{\Gamma_n}=\frac{\omega_n}{\Gamma_{abs}+\Gamma_n^{rad}} \,.
\end{equation}

\begin{figure}[h!]
\begin{center}
\includegraphics[width=8cm]{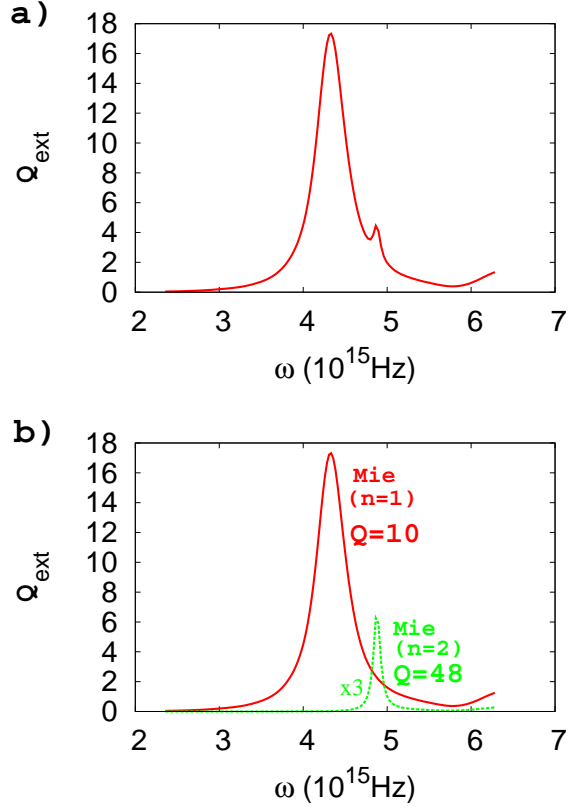}
\caption{\label{fig:Qquad} a) Extinction efficiency of a $R=25~nm$ silver sphere embedded in PMMA, calculated using full Mie expansion. b) Contribution of the dipolar (n=1) and quadrupolar (n=2) mode to the full extinction efficiency. The quality factor is indicated for each mode. Dielectric constant for silver is taken from ref. \cite{Johnson-Christy:1972}.}
\end{center}
\end{figure}

Figure \ref{fig:Qquad} details the quality factor the two first modes of a silver sphere in PMMA using tabulated value for $\epsilon_m$. 
The quadrupolar mode presents a quality factor almost 5 times higher than the dipolar mode since it has limited radiative losses. 
Indeed, quadrupolar mode poorly couples to the far-field. Finally, similar Q values ($Q_n\approx 50$) are obtained for all the higher modes ($n \ge 3$). 
Here again, assuming a Drude metal and in the quasi-static approximation, we qualitatively explains this result. Actually, Q factors of high order modes are absorption loss limited 
($\Gamma_n^{rad} \ll \Gamma_{abs}$, so called dark modes) and tend to $Q_{\infty}=\omega_{\infty}/\Gamma_{abs}$. 
However this overestimates the mode quality factor ($Q_{\infty} \approx 207$) as compared to the value deduced using tabulated value for 
$\epsilon_m$ and exact Mie expansion. 

\subsection{Effective volumes}
Last, we express the {\it total} coupling strength of a dipolar emitter to the spherical metallic particle 
for very short separation distances ($k z_0 \ll 1$) \cite{OptExpGCF:2008,GCFIJMS:2009}:
\begin{equation}
\label{eq:gz}
\frac{\gamma_{tot}^{\perp}}{\gamma_0}\approx\frac{3}{2}\frac{1}{(k z_0)^3}\sum_{n=1}^{\infty}
\frac{(n+1)^2}{z_0^{(2n+1)}}Im(\alpha_n^{eff}) \,.
\end{equation}
So that the coupling strength to the $n^{th}$ mode is easily deduced as 
\begin{eqnarray}
\frac{\gamma_n^{\perp}}{\gamma_0}&\approx&\frac{3}{2}\frac{1}{(k z_0)^3}
\frac{(n+1)^2}{z_0^{(2n+1)}}Im(\alpha_n^{eff}) \,,\\
&\underset{\omega_n}{\sim}&\frac{3}{2}\frac{R^{2n+1}}{(k_n z_0)^3}
\frac{(n+1)^2}{z_0^{(2n+1)}}Q_n \,.
\end{eqnarray}
where we have used approximated expression (\ref{eq:alphaNapprox}) for the $n^{th}$ polarisability. The mode effective volume is then straightforwardly deduced from comparison to the Purcell factor (Eq. \ref{eq:Purcell})
\begin{eqnarray}
V_n^{\perp}=\frac{4\pi z_0^{2n+4}}{(n+1)^2R^{2n+1}} \,.
\end{eqnarray}
We obtain for instance quadrupole effective volume $V_{2}^{\perp}=4\pi z_0^8/(9R^5)=3.2~10^{-4}~\mu m^3=(68~nm)^3$ for an emitter 10 nm away from a 25 nm radius metallic particle in air. 
\begin{figure}[h!]
\begin{center}
\includegraphics[width=6cm,angle=-90]{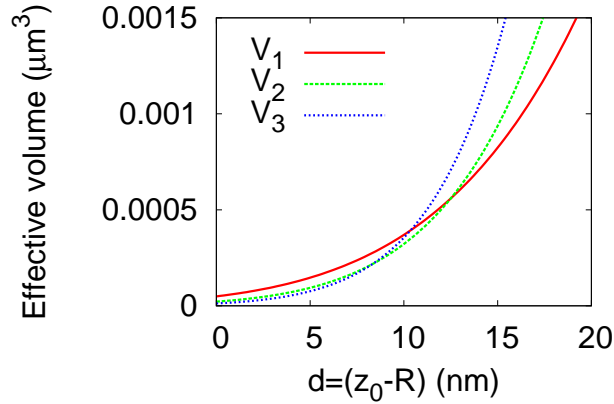}
\caption{\label{fig:Veff}Effective volume of the dipole ($V_1$), quadrupole ($V_2$) and hexapole ($V_3$) modes of a $R=25~nm$ sphere as a function of the emitter-particle distance. The emitter is perpendicular to the particle surface.}
\end{center}
\end{figure}
The calculated effective volumes of the three first modes are shown on figure \ref{fig:Veff}. 
At long distance, the smaller effective volume (most efficient coupling to an emitter) 
is associated to the dipole mode since it presents the largest extension. 
When the emitter-particle distance decreases the coupling strength to the quadrupolar ($d\le 12~nm$), then hexapolar ($d\le 10~nm$) 
mode becomes stronger as revealed by their lower effective volume. 
Most efficient coupling between an emitter and one of the particle modes obviously occurs at contact 
since plasmons mode are confined near the particle surface.

\begin{figure}[h!]
\begin{center}
\includegraphics[width=6cm,angle=-90]{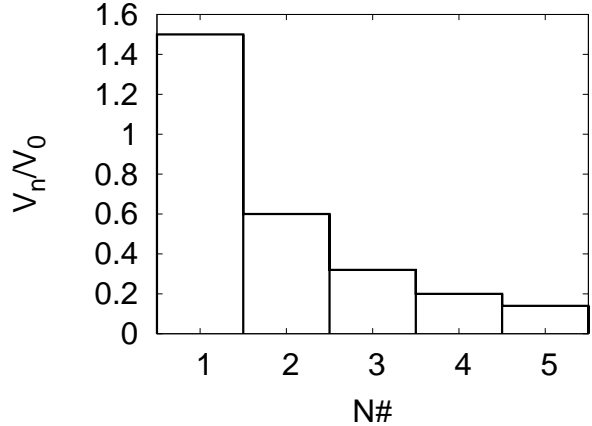}
\caption{\label{fig:Veff0}Normalized effective volume $V_n/V_0$ as a function of mode order.}
\end{center}
\end{figure}

Finally, it is worthwhile to note that mode effective volume is generally defined independently on the particle-emitter distance so that it gives an estimation of the mode extension. This is done by evaluating the maximum effective volume available, and for a random emitter orientation. In case of localized SPP, this is achieved for contact ($z_0=R$). Since the decay rate of a randomly oriented emitter expresses $\gamma=(\gamma^{\perp}+2\gamma^{\parallel})/3$, it comes
\begin{eqnarray}
\frac{1}{{V_n}}&=&\frac{1}{3V_n^{\perp}(R)}+\frac{2}{3V_n^{\parallel}(R)} \,, \\
{V_n}&=&\frac{9}{(2n+1)(n+1)}V_0 \,,
\label{eq:Veff}
\end{eqnarray} 
where $V_0=4\pi R^3/3$ is the metallic sphere volume and $V_n^{\parallel}$ refers to a dipole parallel 
to the sphere surface (see Eq. \ref{eq:Veffx} in appendix, with $\epsilon_B=1$). 
Dipolar and quadrupolar mode volumes are ${V_1}=3/2~V_0$ and ${V_2}=3/5~V_0$, respectively. 
The expression (\ref{eq:Veff}), derived for a Drude metal, {\it quantifies an extremely important property of localized SPPs; 
their effective volume does not depend on the wavelength and is of the order of the particle volume}. 
Metallic nanoparticles therefore support localized modes of strongly subwavelength extension. 
Highest order modes have negligible extension (${V_n} \rightarrow 0$ for $n\rightarrow \infty$, see Fig. \ref{fig:Veff0}). 
As a comparison, photonics cavity modes are generally limited by the diffraction limit so that their effective volume is at best of 
the order of $(\lambda/n_B)^3$ (see ref. \cite{Vahala:2003}). 
Nevertheless, the extremelly reduced SPP volume is achieved at the expense of the mode quality factor. 
It should be noticed that a low quality factor indicates a large cavity resonance FWHM so that emitter-SPP coupling 
can occur on a large spectrum range. 

Note that mode effective volume is generally defined by its energy confinement $V_{eff}=\int \epsilon |{\bf E}({\bf r})|^2 d{\bf r} /max(\epsilon |{\bf E}({\bf r})|^2)$ \cite{Maier:2006}. Sun and coworkers used this definition and obtained \cite{Khurgin-Sun:2009} $V_n=6V_0/(n+1)^2$ that is in agreement with our expression for the dipolar mode volume (${V_1}=1.5V_0$) but leads to slightly different values for other modes (e.g  
${V_2}\approx0.67V_0$ instead of ${V_2}=0.60V_0$). Recently, Koenderink showed that defining the mode volume on the basis of energy density could lead to understimate the Purcell factor near plasmonic nanostructures \cite{Koenderink:2010} (however, he defined the coupling rate to the whole system rather than considering the coupling into a single mode). Oppositely, we adopt here a phenomenological approach where the mode volume is defined so that Purcell factor remains valid. Nevertheless, both methods lead to very similar results for the effective volumes of localized SPPs supported by a nanosphere. 
Since mode volume is generally a simple way to qualitatively characterize the mode extension, expressions derived here or by Sun {\it et al} could be used.

\section{$\beta$ --factor}
\begin{figure}[h!]
\begin{center}
\includegraphics[width=6cm,angle=-90]{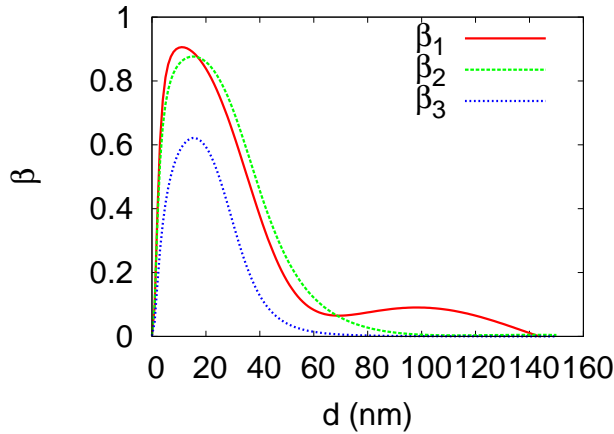}
\caption{\label{fig:beta} Coupling efficiency into the dipolar ($\beta_1$), quadrupolar ($\beta_2$) and hexapolar ($\beta_3$) mode of a $R=25~nm$ silver sphere in PMMA, calculated using exact Mie expansion. The emitter wavelength is assumed to match the considered mode. Therefore $\beta_1$, $\beta_2$ and $\beta_3$ are calculated at $\lambda_1=435~nm$ (dipolar resonance), $\lambda_2=387~nm$ (quadrupolar resonance), and  $\lambda_3=375~nm$ (hexapolar resonance), respectively.}
\end{center}
\end{figure}

Purcell factor quantifies the coupling strength between a quantum emitter and a (plasmon) mode but lacks information 
on the coupling efficiency as compared to all the others emitter relaxation channels. 
For the sake of clarity, it has to be mentionned that coupling strength to one SPP mode corresponds to the total emission decay rate 
induced by this coupling. 
It does not permit to distinguish radiative ($\gamma_{rad}$) and non radiative ($\gamma_{NR}$) coupling into a single mode. 
This could be done by numerically cancelling the imaginary part of the metal dielectric function 
(see also ref. \cite{Barthes-GCF-Bouhelier-Weeber-Dereux:2011} for similar discussion in case of coupling to delocalized SPPs). 
However, one can determine the coupling efficiency into a single mode as compared to all others modes. 
This coupling efficiency, or the so-called $\beta$--factor, is easily estimated in case of a spherical metallic particle 
since all the available channels are taken into account in the Mie expansion. 
Coupling efficiency into $n^{th}$ mode writes
\begin{equation}
\beta_n=\frac{<\gamma_n>}{<\gamma_{tot}>} \,, 
\end{equation}
where $<\gamma>=(\gamma^{\perp}+2\gamma^{\parallel})/3$ is the devay rate of a randomly oriented molecule. $\beta$--factor is represented on Fig. \ref{fig:beta} for the first three modes. A maximum efficiency of $90\%$ can be achieved in the dipolar mode ($d \sim 10~nm$) and $87\%$ into quadrupolar mode ($d\sim 15~nm$). The coupling efficiency into the hexapolar mode is lower ($\sim 60 \%$ around $d\sim15 ~nm$) since it has a very low extension, as indicated in Fig. \ref{fig:Veff0}. 
For very short distances, all the coupling efficiencies drop down to zero since all the higher order modes accumulate in this region, 
opening numerous alternative decay channels. 
Moreover, it is possible to efficiently couple the emitter to either the dipolar or quadrupolar mode, by matching the emitter and mode wavelengths. 
This is of strong importance when designing a SPASER (or plasmon laser) \cite{Bergman-Stockman:2003} 
so that the active mode can be tuned on the dipolar or quadrupolar mode. 
This last {\it spasing} mode would consist of an extremely localized and ultrafast nanosource \cite{Stockman:2010}.

\section{Conclusion}
We explicitly determined the effective volumes for all the SPP modes supported by a metallic nanosphere. 
Their quality factor is also approximated in the quasi-static case or calculated using exact Mie expansion. 
Rather low quality factors ranging from 10 to 100 can be achieved, associated to extremely confined effective volume of nanometric dimensions. 
This results in high $Q/V_{eff}$ ratios indicating efficient coupling strength with a quantum emitter. 
At the opposite to cavity quantum electrodynamics where the coupling strength is obtained on diffraction limited volumes thanks to ultra-high Q factors, plasmonics structures allow efficient coupling on nanometric scales with reasonable Q factors, defining therefore high-bandwidth interaction. Finally, high coupling efficiencies ($80\%-90\%$) to dipolar or quadrupolar mode can be achieved and are of great interest for nanolasers realization.

\section{Acknowledgment}
This work is supported by the Agence
Nationale de la Recherche (ANR) under Grants Plastips (ANR-09-
BLAN-0049-01), Fenopti$\chi$s (ANR-09-
NANO-23) and HYNNA (ANR-10-BLAN-1016). 
This paper was also motivated by the preparation of a lecture for the $2^{nd}$ Summer School
On Plasmonics organized by Nicolas Bonod in october 2011. 

\section{Appendix}
In this appendix, we derive the expression of mode quality factor and effective volume for a spherical particle 
embedded in homogeneous background of optical index $n_B=\sqrt\epsilon_B$. 
For a better description of the metal optical properties, we include the contributions of the bound electrons into the metal dielectric constant 
\begin{equation}
 \epsilon_m=\epsilon_{\infty}-\frac{\omega_p^2}{\omega^2+i\Gamma_{abs} \omega} \,.
\end{equation}
\subsection{Effective polarisability associated to $n^{th}$ SPP mode}
The $n^{th}$ multipole tensor moment of the metallic particle is given by 
\begin{eqnarray}
{\bf p}^{(n)}=\frac {4 \pi \epsilon_0\epsilon_B} {(2n-1)!!} \alpha_n \nabla ^{n-1} {\bf E_0} \,,\\
\alpha_n=\frac{n(\epsilon_m-\epsilon_B)}{n\epsilon_m+(n+1)\epsilon_B}R^{(2n+1)} \,,
\end{eqnarray}
The resonance angular frequency is then $\omega_n=\omega_p \sqrt{n/[n\epsilon_{\infty}+(n+1)\epsilon_B]}$ 
($\omega_{\infty}=\omega_p/ \sqrt{\epsilon_{\infty}+\epsilon_B}$).
Finally, the effective polarisability, including finite size effects write \cite{GCFIJMS:2009}:
\begin{equation}
\alpha_n^{eff}=\left[1 - i~\frac{(n+1)k_B^{2n+1}}{n(2n-1)!!(2n+1)!!}\alpha_n \right ]^{-1}\alpha_n \,.
 \end{equation}
with $k_B=n_Bk$ the wavenumber in the background medium.

Considering a Drude metal, we achieve a simple approximated expression for $\alpha_n^{eff}$ near a resonance
\begin{eqnarray}
\alpha_n^{eff}&\underset{\omega_n}{\sim}&\frac{(2n+1)\epsilon_B}{n\epsilon_{\infty}+(n+1)\epsilon_B}\frac{\omega_n}{2(\omega_n-\omega)-i\Gamma_n} R^{2n+1}\,,\\
\nonumber
\Gamma_n&=&\Gamma_{abs}+\Gamma_n^{rad} \,, \\
\nonumber
\Gamma_n^{rad}&=&\frac{(2n+1)\epsilon_B}{n\epsilon_{\infty}+(n+1)\epsilon_B} \omega_n \frac{(n+1)(k_BR)^{2n+1}}{n(2n-1)!!(2n+1)!!} \,,
\end{eqnarray}

So that the quality factor expression $Q=\omega_n/\Gamma_n$ remains valid but with the 
corrected expressions for resonance frequency $\omega_n$ and total dissipation rate $\Gamma_n$.

\subsection{Effective volumes}
The total coupling strength of a dipolar emitter to the spherical metallic particle, embedded in $n_B$ medium expresses \cite{OptExpGCF:2008,GCFIJMS:2009}:
\begin{eqnarray}
\frac{\gamma_{tot}^{\perp}}{n_B \gamma_0}&\approx&\frac{3}{2}\frac{1}{(k_B z_0)^3}\sum_{n=1}^{\infty}
\frac{(n+1)^2}{z_0^{(2n+1)}}Im(\alpha_n^{eff}) \,, \\
\label{eq:gx}
\frac{\gamma_{tot}^{\parallel}}{n_B \gamma_0}&\approx&\frac{3}{4}\frac{1}{(k_B z_0)^3}\sum_{n=1}^{\infty}
\frac{n(n+1)}{z_0^{(2n+1)}}Im(\alpha_n^{eff}) \,.
\end{eqnarray}
The coupling strength to the $n^{th}$ mode is  
\begin{eqnarray}
\frac{\gamma_n^{\perp}}{n_B\gamma_0}&\approx&\frac{3}{2}\frac{1}{(k_B z_0)^3}
\frac{(n+1)^2}{z_0^{(2n+1)}}Im(\alpha_n^{eff}) \,,\\
\nonumber
&\underset{\omega_n}{\sim}&\frac{3}{4\pi^2}\left(\frac{\lambda}{n_B}\right)^3\frac{R^{2n+1}}{4\pi z_0^3}\frac{(2n+1)\epsilon_B}{n\epsilon_{\infty}+(n+1)\epsilon_B}
\frac{(n+1)^2}{z_0^{(2n+1)}}Q_n \,.\\
\frac{\gamma_n^{\parallel}}{n_B\gamma_0}&\approx&\frac{3}{4}\frac{1}{(k_B z_0)^3}
\frac{n(n+1)}{z_0^{(2n+1)}}Im(\alpha_n^{eff}) \,, \\
\nonumber
&\underset{\omega_n}{\sim}&\frac{3}{4\pi^2}\left(\frac{\lambda}{n_B}\right)^3\frac{R^{2n+1}}{8\pi z_0^3}\frac{(2n+1)\epsilon_B}{n\epsilon_{\infty}+(n+1)\epsilon_B}
\frac{n(n+1)}{z_0^{(2n+1)}}Q_n \,,
\end{eqnarray}
and the mode effective volume is deduced from comparison to the Purcell factor (Eq. \ref{eq:Purcell})
\begin{eqnarray}
V_n^{\perp}=\frac{n\epsilon_{\infty}+(n+1)\epsilon_B}{(2n+1)\epsilon_B}\frac{4\pi z_0^{2n+4}}{(n+1)^2R^{2n+1}} \,,\\
V_n^{\parallel}=\frac{n\epsilon_{\infty}+(n+1)\epsilon_B}{(2n+1)\epsilon_B}\frac{8 \pi z_0^{2n+4}}{n(n+1)R^{2n+1}} \,.
\label{eq:Veffx}
\end{eqnarray}
Finally, we define the mode volume as 
\begin{eqnarray}
\frac{1}{{V_n}}&=&\frac{1}{3V_n^{\perp}(R)}+\frac{2}{3V_n^{\parallel}(R)} \,, \\
{V_n}&=&\frac{n\epsilon_{\infty}+(n+1)\epsilon_B}{(2n+1)\epsilon_B}\frac{9}{(2n+1)(n+1)}V_0 \,.
\end{eqnarray} 
All the expressions obtained in this appendix reduce to the simple analytical case discussed in the main text for 
$\epsilon_{\infty}=1$ (bound electrons contribution neglected) and $\epsilon_{B}=1$ (background medium is air).
%


\begin{thebibliography}{34}
\bibitem{Notation}{for the sake of clarity, we use lowercase notation $\gamma$ for the emitter decay rates 
and uppercase $\Gamma$ for the quantities associated to the metallic particle.}
\end{thebibliography}
\end{document}